\documentclass[lettersize,journal]{IEEEtran}
\usepackage{amsmath,amsfonts}
\usepackage{algorithmic}
\usepackage{algorithm}
\usepackage{array}
\usepackage[caption=false,font=normalsize,labelfont=sf,textfont=sf]{subfig}
\usepackage{textcomp}
\usepackage{stfloats}
\usepackage{url}
\usepackage{verbatim}
\usepackage{graphicx}
\usepackage{cite}
\usepackage{color}
\usepackage{diagbox}
\usepackage{flushend}
\begin{document}

\title{Network Intrusion Detection with Edge-Directed Graph Multi-Head Attention Networks}

\author{Xiang~Li, Jing~Zhang,~\IEEEmembership{Senior~Member,~IEEE,} Yali Yuan, Cangqi Zhou,~\IEEEmembership{Member,~IEEE,}
	\thanks{X. Li, J. Zhang and Y. Yuan are with the School of Cyber Science and Engineering, Southeast University, Nanjing 211189, China. J. Zhang is also with the Engineering Research Center of Blockchain Application, Supervision and Management (Southeast University), Ministry of Education, Nanjing 211189, China. J. Zhang is the corresponding author of this paper. E-mail: jingz@seu.edu.cn.}
	\thanks{C. Zhou is with the School of Computer Science and Engineering, Nanjing University of Science and Technology, Nanjing 210094, China.}
\thanks{Manuscript received December 1, 2023; revised January 1, 2024.}}

\markboth{Journal of \LaTeX\ Class Files,~Vol.~99, No.~10, October~2023}%
{Li \MakeLowercase{\textit{et al.}}: Network Intrusion Detection with Edge Graph Attention Networks}

\IEEEpubid{0000--0000/00\$00.00~\copyright~2023 IEEE}

\maketitle

\begin{abstract}
A network intrusion usually involves a number of network locations. Data flow (including the data generated by intrusion behaviors) among these locations (usually represented by IP addresses) naturally forms a graph. Thus, graph neural networks (GNNs) have been used in the construction of intrusion detection models in recent years since they have an excellent ability to capture graph topological features of intrusion data flow. However, existing GNN models treat node mean aggregation equally in node information aggregation. In reality, the correlations of nodes and their neighbors as well as the linked edges are different. Assigning higher weights to nodes and edges with high similarity can highlight the correlation among them, which will enhance the accuracy and expressiveness of the model. To this end, this paper proposes novel Edge-Directed Graph Multi-Head Attention Networks (EDGMAT) for network intrusion detection. The proposed EDGMAT model introduces a multi-head attention mechanism into the intrusion detection model. Additional weight learning is realized through the combination of a multi-head attention mechanism and edge features. Weighted aggregation makes better use of the relationship between different network traffic data. Experimental results on four recent NIDS benchmark datasets show that the performance of EDGMAT in terms of weighted F1-Score is significantly better than that of four state-of-the-art models in multi-class detection tasks.
\end{abstract}


\section{Introduction}
\IEEEPARstart{S}{ecurity} and reliability have always been one of critical concerns in computer networks and systems. As network structure becomes larger and more complex, computer networks also face a variety of threats and attacks, such as fraud, viruses, worms, spyware, and malware. Due to the increase in network complexity, network systems are subject to many security vulnerabilities, including intrusions. Therefore, in order to effectively defend against different types of attacks, it is imperative to develop an accurate intrusion detection system (IDS) \cite{Liao2013IntrusionDS}. There exist two main types of IDS, signature-based detection \cite{McHugh2000TestingID} and anomaly-based detection \cite{GarcaTeodoro2009AnomalybasedNI}. The signature-based intrusion detection methods rely on the pre-constructed network attack signature database. They match the monitored network traffic with the signature patterns to detect attack behaviors. Therefore, the signature-based methods are difficult to deal with unknown attacks and are helpless against some specified network intrusions such as zero-day attacks. Anomaly-based intrusion detection methods utilize the deviation between current traffic and normal traffic model and classify traffic according to the differences between features. Therefore, such methods are more suitable for dealing with dynamic and unknown intrusion types, which play an increasingly important role in network security.

It is known that machine learning is a fundamental technology for building anomaly-based intrusion detection systems. The early intrusion detection models were usually built with unitary learning models such as SVM \cite{Zhang2005ApplicationOO,lvarezCidFuentes2020AdaptivePA}, decision trees \cite{Sindhu2012DecisionTB}, etc. To improve the generalization and expression ability of models, more advanced methods resort to ensemble learning algorithms such as XGBoost \cite{lawal2020anomaly} and random forest \cite{kumar2021ensemble}. However, on the one hand, these machine learning-based methods heavily rely on feature engineering, and on the other hand, they lack the ability to deal with large-scale data \cite{ahmad2021network}. In recent years, the development of deep learning, which can learn feature representations from massive raw data, has provided a good opportunity to address these issues. Classical deep neural networks such as CNN\cite{Elsayed2021ANH}, LSTM \cite{he2019novel}, and RNN \cite{yin2017deep} have already been used to build high-performance IDS. 
\IEEEpubidadjcol

The intrusion detection models based on traditional neural network structures (such as CNN, LSTM, RNN, etc.) use statistical characteristics and protocol fields in network flow to build network traffic classifiers. One common weakness of these models is the lack of network topology within the models, which fails to consider the interrelationships between network flow records and ignored the features conveyed by the topology information represented by IP addresses and port numbers. Moreover, many network flow data are nonlinear and high-dimensional, which means that some features in the data are correlated and others are redundant, making feature extraction more difficult. To this end, a more advanced deep learning approach, graph neural networks (GNN), has begun to enter the field of IDS. GNNs can encode non-Euclidean graph data for representation learning and subsequent classification prediction. A communication network is inherently graph-structured, including nodes and edges. The IP addresses and port numbers are combined as sockets to map to the nodes uniquely, while the flow data flowing through the network is mapped as directed edges. It is natural to model the entire network flow using GNNs. In 2021, Lo et al. \cite{Lo2021EGraphSAGEAG} proposed an E-GraphSAGE method, which utilizes graph convolutional neural networks (GCNs). In 2022, Zhang et al. \cite{zhang2022intrusion} proposed an intrusion detection model based on GID.


Neighborhood aggregation is a fundamental operation of GNNs. Existing GNN-based models such as E-GraphSAGE \cite{Lo2021EGraphSAGEAG}) treat the mean aggregation of nodes in the neighborhood equally. However, in reality, the correlations of nodes and their neighbors as well as the linked edges are different. Assigning higher weights to nodes and edges with high similarity can highlight the correlation among them, which may improve the expressiveness of the model. Intuitively, dynamical aggregation of nodes and edges can be achieved by Graph Attention Networks (GAT) \cite{Velickovic2017GraphAN}, an attention-mechanism-based GCN. However, directly applying GAT to intrusion detection is infeasible. First, GAT only aggregates nodes. It does not aggregate weighted edges with features. The features and weights on edges are especially important to network flow. Second, GAT was designed for undirected graphs, while in our scenario, the directions of network flow also matter. Thus, in this study, we propose novel Edge-Directed Graph Multi-Head Attention Networks (EDGMAT) for network intrusion detection. The proposed EGAT model employs a multi-head attention mechanism to aggregate nodes and edges with different weights. It can better explore correlation among nodes and edges as well as make better use of the whole network's topological information. To sum up, the contributions of this paper are three-fold:
\begin{itemize}
	\item[$\bullet$] We propose a novel EDGMAT model for network intrusion detection, which utilizes a multi-head attention mechanism to aggregate nodes and edges. The model makes full use of node features and edge features and better explores correlations among nodes and edges by assigning different weights.
	\item[$\bullet$] The proposed multi-head attention mechanism in EDGMAT is specially designed for intrusion detection, which can exploit network topological features such as edge directions and weights.
	\item[$\bullet$] The proposed EDGMAT model is extensively evaluated on four benchmark network intrusion detection datasets. Experimental results consistently demonstrate the classification performance of EDGMAT is significantly better than that of the state-of-the-art methods.
\end{itemize}

The remainder of the paper is organized as follows: Section~\ref{sec:rw} briefly reviews the network intrusion detection methods. Section~\ref{sec:pk} presents some prerequisite knowledge used in this study. Section~\ref{sec:mtd} presents the proposed EDGMAT model in detail. Section~\ref{sec:exp} presents the experiments and discusses the results. Section~\ref{sec:con} concludes the paper and discusses future work.

\section{Related Work}\label{sec:rw}
Machine learning technology has been widely used to build efficient and accurate IDS in recent years. Many machine learning models can be applied to the classification of network traffic. Churcher et al.\cite{churcher2021experimental} investigated the effectiveness of various machine learning algorithms, including k-nearest neighbor (KNN), decision trees (DT), support vector machines (SVM), naive Bayes (NB), random forest (RF), artificial neural network (ANN), and logistic regression (LR), for network intrusion detection. Their experimental results show that KNN has the best performance. To evaluate machine learning models across datasets and test their ability to generalize to different network environments and attack scenarios, Sarhan et al. \cite{sarhan2021netflow} provided multiple datasets with Netflow-based feature sets. These data sets were evaluated using the Extra Trees algorithm. Lawal et al. \cite{lawal2020anomaly} implemented a hybrid network intrusion detection system based on features and behaviors, using a database based on signatures and blacklisted sources. Their anomaly detection model based on the Extreme Gradient Lift algorithm (XGBoost) can achieve fast and accurate attack detection. Li et al. \cite{li2021sustainable} proposed a sustainable incremental ensemble learning model, which adaptively allocated the decision weights of individual classifiers using multi-class regression models. The model adopts an iterative update strategy, where it adds the parameters and decision results of the historical model to the training process of a new ensemble model.

As a new development of machine learning technology, deep learning mainly aims to solve the high-level abstract concept learning and realizes the low-level feature extraction by establishing deep neural networks\cite{jasim2022survey}. Farahnakian et al. \cite{farahnakian2018deep} introduced deep learning to IDS, which employed a deep autoencoder (DAE) as a deep encoding technique. To avoid over-fitting and local optima, the DAE technology proposed in the article adopts a greedy layered approach for training. Shone et al. \cite{shone2018deep} proposed an unsupervised feature learning method with asymmetric deep autoencoder (NDAE). The authors also built a new deep learning classification model using stacked NDAE and validated it on some basic datasets. In addition, network intrusion detection models based on RNN\cite{li2018using}, DNN\cite{rm2020effective} and CNN\cite{zhang2019improved} have also been proposed. With the help of the automatic feature extraction function of the models, the intrusion detection system based on deep learning eliminates the feature extraction project, and more data also improves the prediction accuracy\cite{gottwalt2019corrcorr}.

Intrusion detection based on graph neural networks is a new direction in network intrusion and anomaly detection. Since GNNs have the ability to capture hidden network topology and extend it to invisible topology \cite{chang2021graph}, the GNN-based intrusion detection method integrates graph topology and network flow features to achieve more accurate judgment and higher generalization than traditional neural networks. Xiao \cite{xiao2020towards} et al. proposed an anomaly detection model for graph embedding methods based on network flow representations. The model converts network traffic into first-order and second-order graphs. The first-order graph learns latent information in the network from the perspective of a single host, while the second-order graph learns latent features from a global perspective. It uses the extracted graph embedding and eigenvalues to train a random forest classifier using the transductive method \cite{hamilton2017inductive}, and successfully applies network topology features to intrusion detection. Lo et al. \cite{Lo2021EGraphSAGEAG} proposed an E-GraphSAGE model based on GCN for intrusion detection in IoT. The model can collect information in the graph to update node feature representations and allows edge features to be considered. It first constitutes the network flow topology through the source nodes and the target nodes, then performs convolution operation on the remaining flow features as edge features, and realizes the feature update through the sampling and aggregation operation of adjacent node information. Zhou et al. \cite{zhou2020automating} proposed a GNN model for end-to-end botnet detection. Their method only considers the topology of the network, omits the edge features, and set the node features as features containing only constant vectors. It enables the model to aggregate features across the graph without considering the influence of unique node attributes.  It significantly improves botnet detection compared to results obtained by logistic regression and an existing botnet detection tool BotGrep \cite{nagaraja2010botgrep}. Zhang et al.\ cite{zhang2022intrusion} proposed a reconfigurable neural network-based intrusion detection system for IoT, namely Graph Intrusion Detection (GID). It realizes a network embedding feature representation method, which reduces the impact of network structure inaccuracy on the model by designing a regularized network constructor, and finally uses the network embedding representation weight and network constructor to train together. It has good classification ability for high-dimensional, redundant but unbalanced rare labeled data.

Graph Attention Network (GAT) has been widely used in graph data processing. The key of the GAT model is to integrate the attention mechanism into propagation and node aggregation, and the learning result is more stable when using a multi-head attention mechanism. There are also relevant studies on the attention network combining edge features \cite{wang2021egat}, which assigns different weights to different nodes and edges by calculating the similarity between nodes and makes decisions on the most important neighbors. Heterogeneous Edge Feature Graph Attention Networks (HeEdgeGAT) \cite{monninger2023scene} encodes a variety of traffic scenarios through heterogeneous graphs, and uses heterogeneous graph neural network encoders and task-specific decoders to make reasoning. Chang \cite{chang2021graph} proposed an E-ResGAT model. It uses residual learning to preserve the original information by adding residual connections to the resulting edge embeddings to improve the performance of predicting a few categories. In recent years, there is few study on intrusion detection based on graph attention networks. Thus, we proposed a model based on the edge graph attention mechanism, which employs edge features and a multi-head attention mechanism and shows good performance in multi-classification tasks.

\section{Prerequisite Knowledge}\label{sec:pk}
In this section, we will introduce some prerequisite knowledge used in our study.

\subsection{Graph Attention Networks}
Graph attention networks (GATs)\cite{Velickovic2017GraphAN} are some kind of graph convolutional neural networks (GCNs) based on attention mechanisms. Different from the traditional GCN model, the GAT model utilizes the graph attention network to dynamically aggregate node features, utilizes the relationship between nodes and features to calculate the similarity as the feature weight, and then updates the feature representation of the central node by weighted aggregation. The attention coefficient in GAT is obtained by learning, and the weights between different nodes can be learned adaptively, so as to better capture the interrelationships between nodes. GAT acts in a graphic format to obtain structural information by modeling a set of objects and their relationships. In the case of network intrusion detection, IP addresses are modeled as graph nodes and network flow information between hosts is modeled as graph edges. 

Therefore, we can define a graph $\mathcal{G}(\mathcal{V}, \mathcal{E})$, where $\mathcal{V}$ is the set of nodes and $\mathcal{E}$ is the set of edges. The input of the GAT structure is defined as a set of node features $\mathbf{H}=\left\{\vec{h}_1, \vec{h}_2, \ldots, \vec{h}_N\right\}, \vec{h}_i \in \mathbb{R}^F$, and the result output is a set of updated high-dimensional features $\mathbf{H}^{\prime}=\left\{\vec{h}_1^{\prime}, \vec{h}_2^{\prime} \ldots, \vec{h}_N^{\prime}\right\}, \vec{h}_i^{\prime} \in \mathbb{R}^{F^{\prime}}$. The model uses a learnable weight matrix $\mathbf{W}_H \in \mathbb{R}^{F \times F^{\prime}}$ and an attention mechanism \textit{a} to calculate the node attention coefficient $e_{i j}=a\left(\mathbf{W} \vec{h}_i, \mathbf{W} \vec{h}_j\right)$ which represents the importance of the features of node \textit{j} to node \textit{i}. For the easy comparison of the coefficients between different nodes, it is necessary to normalize the attention coefficient as follows:
\begin{equation}
	\alpha_{i j}=\operatorname{softmax}_j\left(e_{i j}\right)=\frac{\exp \left(e_{i j}\right)}{\sum_{k \in \mathcal{N}_i} \exp \left(e_{i k}\right)},
\end{equation}
where $\textit{N}_{\textit{i}}$ is a neighborhood of node \textit{i} in the graph. Consequently, the normalized attention coefficient is used to calculate a linear combination of the corresponding features as the final output for each node.

\subsection{Edge-Featured Graph Attention Networks}
Edge-featured graph attention networks (EGATs)\cite{wang2021egat} are graph neural networks based on the annotation mechanism, which is specifically used to process graph data with edge features. In order to make effective use of edge features, the EGAT model enhances the original attention mechanism and regards edge information as an important factor in the calculation of attention weights. EGATs accept a set of node features $\mathbf{H}=\left\{\vec{h}_1, \vec{h}_2, \ldots, \vec{h}_N\right\}, \vec{h}_i \in \mathbb{R}^F$ and a set of edge features $\mathbf{E}=\left\{\vec{e}_1, \vec{e}_2 \ldots, \vec{e}_M\right\}, \vec{e}_p \in \mathbb{R}^{F_E}$, calculate the similarity between each node and its neighbor nodes, measure the degree of association between nodes. All node and edge features will be aggregated in the final layer so that the model can learn the necessary features from various scales to facilitate classification. There are two important attention components in the EGAT model, i.e., node attention block and edge attention block.

\subsubsection{Node Attention Block}
For each node in the graph, the algorithm calculates the similarity between the node and the neighbor node, which is used to measure the correlation degree between the nodes, and calculates the attention coefficient of the neighbor node by integrating the edge feature attention mechanism as follows:
\begin{equation}
	\alpha_{i j}=\frac{\exp \left(\operatorname{LeakyReLU}\left(\overrightarrow{\mathbf{a}}^T\left[\vec{h}_i\left\|\vec{h}_j\right\| \vec{e}_{i j}\right]\right)\right)}{\sum_{k \in \mathcal{N}_i} \exp \left(\operatorname{LeakyReLU}\left(\overrightarrow{\mathbf{a}}^T\left[\vec{h}_i\left\|\vec{h}_k\right\| \vec{e}_{i k}\right]\right)\right)}.
\end{equation}

After obtaining the attention weight for each neighborhood, the normalization process is performed, followed by a weighted summation of node features within these neighborhoods. The above process results in the output of the node attention module, which represents the updated node features. Mathematically, the updated node features can be expressed as follows:
\begin{equation}
	\vec{h}_i^{\prime}=\sigma\left(\sum_{j \in \mathcal{N}_t} \alpha_{i j} \vec{h}_j\right).
\end{equation}

\subsubsection{Edge Attention Block}
Node features update themselves periodically in the node attention module to obtain higher-level features. Thus, it is unreasonable to use the original edge features in the calculation of characteristic polymerization. That is, we also need higher-level edge features. In addition, higher-level edge features are also required in the EGAT layer to maintain a balance of importance between nodes and edges. Therefore, an edge attention module is required in the model, which again takes in a set of node features and a set of edge features and produces a new set of edge features. In order to update the edge features, EGAT creates a new graph, using the nodes and edges in the original graph as the edges and nodes in the new graph, and generates the new edge features through the same process as in the node module. The attention coefficient between the edges can be expressed as follows:
\begin{equation}
	\beta_{p q}=\frac{\exp \left(\operatorname{LeakyReLU}\left(\vec{b}^T\left[\vec{e}_p\left\|\vec{e}_q\right\| \vec{h}_{p q}\right]\right)\right)}{\sum_{k \in \mathcal{N}_p} \exp \left(\operatorname{LeakyReLU}\left(\overrightarrow{\mathbf{b}}^T\left[\vec{e}_p\left\|\vec{e}_k\right\| \vec{h}_{p k}\right]\right)\right)}.
\end{equation}
After aggregation, the new edge features are represented as follows:
\begin{equation}
	\vec{e}_p^{\prime}=\sigma\left(\sum_{q \in \mathcal{N}_p} \beta_{p q} \vec{e}_q\right).
\end{equation}

The EGAT model successfully merges edges into entities equivalent to nodes, points out the different weights that the graph treats data, and realizes the process of spontaneous learning through the attention mechanism. The model implements a more efficient and accurate graph structure feature aggregation algorithm, which effectively improves the node classification performance of the model in edge feature-sensitive datasets. The model has been applied and performed well in tasks such as wind farm power prediction, knowledge graph prediction in the traffic environment, and sentence sentiment analysis. In the field of network security, the model has also been successfully applied to the detection of phishing accounts in the blockchain platform\cite{zhou2023detecting}, showing performance better than the current Graph2Vec\cite{kim2022graph} and DeepWalk models\cite{tao2022structural}.
\section{The Proposed Method}\label{sec:mtd}
In this section, we first present the motivation for proposing a novel model. Then, we present the structure of the proposed model, followed by its technical details. Finally, we summarize the proposed method as an algorithm.

\subsection{Motivations}
Exploring correlations among nodes and edges is critical in neighborhood aggregation of graph learning. Assigning greater weights to nodes and edges with high similarity can highlight the correlation among them. Therefore, it is a straightforward idea that we can incorporate Graph Attention Networks (GAT) \cite{Velickovic2017GraphAN} into the network intrusion detection system to enhance weight learning. However, as we know, one weakness of GAT is that it does not include edge features when performing node embedding. In the context of the network flow graph, edge features (e.g., source and destination IP addresses, ports, traffic volume, etc.) play a crucial role in distinguishing malicious traffic. Therefore, it will be an intuitively better solution to introduce edge features into the neighborhood aggregation and updating of node features while ensuring their appropriate adjustment. Consequently, an off-the-shelf model, namely  Edge-Featured Graph Attention Network (EGAT) \cite{wang2021egat}, may have a good potential to be applied in the IDS. Nevertheless, in our study, we noticed that EGAT does not account for the directed graph structure inherent in the network flow environment, where traffic flows from one node to another. Traffic direction is undoubtedly an important factor that can help determine whether a data flow is an intrusion or not. We need a novel viable solution to address these challenges.

\subsection{Network Structure of the Proposed Model}
\begin{figure}[!t]
	\centering
	\includegraphics[width=3.5in]{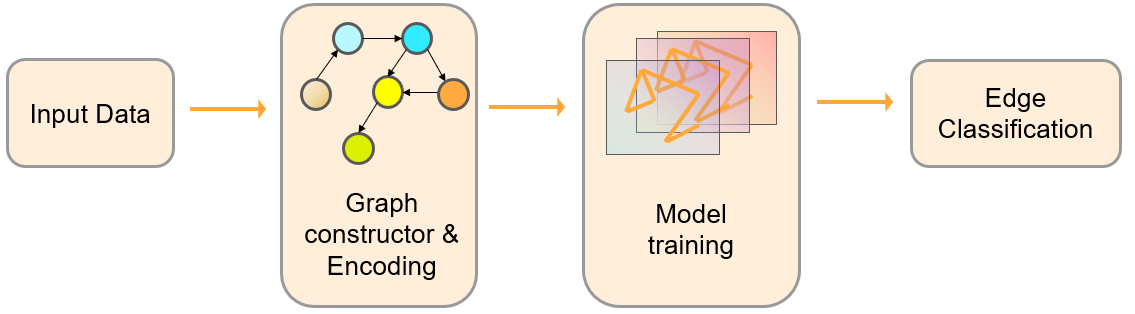}
	\caption{Sketch of our proposed model.}
	\label{fig1}
\end{figure}
In this section, we introduce the framework of our proposed model. The framework encodes the original network flow data and generates a network flow graph for model training. The network flows are then classified based on the edge embeddings obtained by the neural network. Figure 1 illustrates the structure of our proposed framework. The graph building module generates the corresponding network flow graph based on the data flow information. These network flow records typically include fields that identify communication, such as source and destination addresses. Additionally, the records provide further details about the flow, such as the number of packets, bytes, and duration.

Our approach draws inspiration from previous work \cite{xiao2020towards} and employs four features to define the edges of the graph: source IP address, source port number, destination IP address, and destination port number. The first two features form a tuple representing the source node, while the last two features represent the destination node of the flow.

To construct the graph, we utilize a multiway approach that incorporates the topological direction of the network flow from the source IP address and port number to the destination IP address and port number. By connecting adjacent nodes, we create a single directed graph. This approach differs from the previous method \cite{Lo2021EGraphSAGEAG} of establishing positive and negative connections between adjacent nodes to determine edge directionality. Our approach enhances our understanding of topology and simplifies the complexity of the graph. Furthermore, by defining the direction of edges in the topology construction process , we reduce the number of edges in the graph and increase training efficiency.

In addition, when applying a graph neural network to network intrusion detection, it is essential to consider the topological effects of both transductive and inductive settings. In the transductive setting, we incorporate all data into a graph for training, utilizing a fixed network topology for network intrusion detection. Conversely, in the inductive setting, we dynamically build the network topology based on the given network flow data. The advantages and disadvantages of both settings have been extensively discussed in prior research\cite{bianchini2016comparative}. In this article, we assess the performance of both manners within our proposed model and network topology.

After constructing the graph, the next step is to train the model on the graph using the provided data set. The graph features are updated through the process of node aggregation, which aggregates the information from neighboring nodes to obtain new edge embeddings. These updated edge embeddings are then used for the final edge classification. During the training process, the model aims to minimize the loss function, which measures the discrepancy between the predicted edge classifications and the ground truth. The parameters in the model will be updated iteratively to improve the classification performance. Once the model has been trained, we can use it to classify network flow records in the test data set. These records are converted into graph representations, and the model calculates new edge embeddings for each graph. The final classification results are obtained by applying the softmax layer, which assigns probabilities to each class.

\subsection{Technical Details}
\begin{figure}[!t]
	\centering
	\includegraphics[width=3.5in]{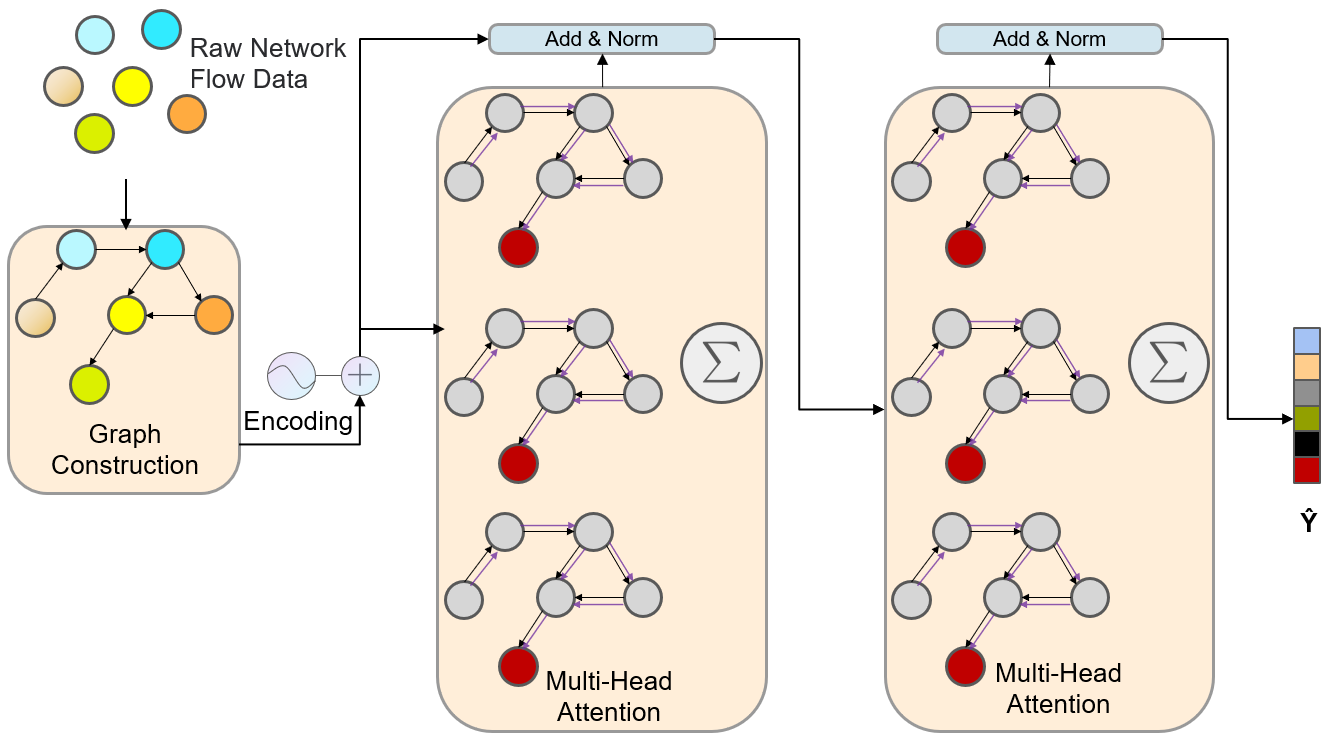}
	\caption{Schematic representation of our proposed method for intrusiondetection.The black arrow indicates the direction of network traffic transmission, and the purple arrow indicates the direction of neighborhood aggregation.}
	\label{fig2}
\end{figure}

In the proposed model, we use the digraph $\mathcal{G}(\mathcal{V}, \mathcal{E})$ to define a network topology composed of intrusion detection datasets. Each node $v_i \in \mathcal{V}$ has a characteristic vector $h_i$, and the edeg $e_{j, i}=\left(v_j, v_i\right) \in \mathcal{E}$ between the source node $v_j$ and the target node $v_i$ has a characteristic vector $\textbf{e}_{j, i}$. $\mathcal{N}(i)$ represents the set of neighbor node of the node \textit{i}, and \textit{T} represents the number of layers of the convolution layer.

Based on the attention method in Section~\ref{sec:pk}, we construct the edge graph attention network convolution (EdgeGATconv) layer as the main part of our model. EdgeGATConv layer is mainly used for edge classification or graph-level tasks in graph neural networks. We use this layer to identify malicious traffic in network intrusion detection tasks. Figure \ref{fig2} shows the main structure of the EdgeGATConv layer we built, and then we introduce its construction ideas in detail.

To better understand the forward calculation process, we define the essential parameters involved, including the graph data required by the convolutional layer, the input node features, input edge features, output node features, output edge features, and the number of heads for the multi-head attention mechanism. First, a linear transformation layer $W_{\mathrm{n}}^k \cdot h_i$ is constructed based on the input feature dimension. This maps the input feature to the desired output feature dimension of the multi-head attention. Similarly, a linear transformation $W_{\mathrm{e}}^k \cdot \mathbf{e}_{j, i}$ is applied to the edge features, mapping them to the output feature dimensions of the multi-head attention. Subsequently, the multi-head attention mechanism $\mathbf{a}^{k^T}$ is employed to calculate the attention weights $\alpha_{j, i}^k$ between source node features and target node features. By utilizing these node features and their respective attention weights, the node's attention representation is obtained. The attention representations of both the source and target nodes are then combined, and a weighted summation of the edge features is performed to update the edge features
\begin{equation}
\textbf{e}_{ij}^{\prime} = a_{ij} \cdot [W \cdot h_j \, || \, W \cdot \textbf{e}_{ij}]. 
\end{equation}
The weighted edge features are then combined with the source node features, thereby aggregating the new features $h_i^{\prime}$ of the destination node. This aggregation process involves the addition of residuals and bias to the aggregated node features. Finally, the output features undergo a non-linear transformation using an optional activation function. According to the definition of the convolution layer, the attention coefficient between nodes can be obtained by the  equation as follows:
\begin{equation}
	\begin{aligned}
		\alpha_{j, i}^k= & \operatorname{softmax}_i(\operatorname{LeakyReLU}( \\
		& \left.\left.\mathbf{a}^{k^T}\left[W_{\mathrm{n}}^k \cdot h_i\left\|W_{\mathrm{n}}^k \cdot h_j\right\| W_{\mathrm{e}}^k \cdot \mathbf{e}_{j, i}\right]\right)\right),
	\end{aligned}
\end{equation}
where \textit{W} is used to represent a learnable weight matrix. The matrix can transform features to update node features, adjacent node features, and edge features. \textit{K} represents the number of heads in the multi-head attention mechanism. After obtaining the attention coefficient, we define the node update as follows:
\begin{equation}
	\begin{aligned}
		h_i^{\prime}= 	& W_{\mathrm{s}} \cdot h_i+ \\
		& \|_{k=1}^K\left(\sum_{j \in \mathcal{N}\left(v_i\right)} \alpha_{j, i}^k\left(W_{\mathrm{n}}^k \cdot h_j+W_{\mathrm{e}}^k \cdot \mathbf{e}_{j, i}\right)\right).
	\end{aligned}
\end{equation}
To better propagate errors and avoid excessive smoothing, we add residual connections $W_{\mathrm{s}} \cdot h_i$ during the aggregation of graph attention network nodes. Residual joins add the output of the previous layer directly to the output of the current layer, thus providing a path around the nonlinear transformation. In this way, the network can learn to retain important information after the information is compressed or stretched \cite{he2020resnet}, while it also mitigates the problem of gradient disappearance or explosion.

After passing through multiple graph convolution layers, the model finally outputs updated node embeddings and edge embeddings, which are not the classification results desired in this paper. For the specific task of network intrusion detection, this study uses a softmax layer as the decoder and applies it to the multi-class edge classification coding considered in the experimental section to obtain the final classification output and compare it with the real label to calculate the performance indicators of the classification model.

\subsection{Algorithm}
Algorithm~\ref{alg1} presents the training process of the proposed Edge-Directed Graph Multi-Head Attension Network (EDGMAT) model.
\begin{algorithm}[H]
	\caption{The EDGMAT Model Training}
	\begin{algorithmic}[1]
		
		\REQUIRE ~~\\
		Raw netflow data \textbf{X} \\
		Heads of attention \textit{K}
		\ENSURE 
		Predicted label $\hat{\mathbf{Y}}$ \\
		\STATE $t \leftarrow 1$
		\STATE 	$\mathcal{G}(\mathcal{V}, \mathcal{E}) \leftarrow \textbf{X} $
		\STATE Generate node features $h_i,v_i \in \mathcal{V} \leftarrow \textbf{X},\mathcal{G}(\mathcal{V}, \mathcal{E})$
		\STATE Generate edge features $\textbf{e}_{i j}, \forall i j \in \mathcal{E}\leftarrow \textbf{X},\mathcal{G}(\mathcal{V}, \mathcal{E})$

		\FOR{$t \leq T$}
		\FOR{$v_i \in \mathcal{V}$}
		\STATE  Linearly transform node features $h_i$
		\STATE  Linearly transform edge features $\textbf{e}_{ij}$
		\STATE $ \beta_{ij} = \text{LeakyReLU} \left( (\mathbf{a}^T) \cdot [W \cdot h_i \, || \, W \cdot h_j \, || \, W \cdot \textbf{e}_{ij}] \right) $ $\rhd$ Calculate the attention coefficient of the node
		\STATE $\alpha_{ij} = \text{softmax}(\beta_{ij}) $ Normalize the attention coefficient 
		\STATE $  h_i^{\prime} =\|_{k=1}^K\left(\sum_{j \in N_i} (a_{ij} \cdot [W \cdot h_j \, || \, W \cdot \textbf{e}_{ij}])\right) $ Aggregate to get new edge features
		\STATE $  \textbf{e}_{ij}^{\prime} = a_{ij} \cdot [W \cdot h_j \, || \, W \cdot \textbf{e}_{ij}] $
		\STATE $h_i^{\prime} = h_i^{\prime} + h_i$ Get residual connection
		\STATE $h_i^{\prime} = \text{activation}(h_i^{\prime})$
		\ENDFOR
		\STATE $\hat{\mathbf{Y}} \leftarrow \textbf{e}_{ij}^{\prime}$ Generate predict label
		\STATE $L^t \leftarrow l\left(\hat{\mathbf{Y}}_{\text {train }}, \textbf{Y}_{\text {train }}\right)$ Calculate the predicting loss of training data
		\STATE Back-propagate loss to update model parameters
		\STATE $t \leftarrow t+1$
		\ENDFOR
	
	\end{algorithmic}
	\label{alg1}
\end{algorithm}

In the initial four lines of the algorithm, we specify the graph structure and feature structure, followed by a comprehensive description of the algorithm definition. Lines 7 and 8 define the output dimensions for the features and linear transformations within the convolution layer. Subsequently, the attention coefficient is calculated in the subsequent two lines. Lines 11 and 12 encompass the updates made to both node and edge features, along with the addition of residual connections. During the training process, the model iteratively adjusts itself to reach a locally optimal solution by calculating the loss function using the predicted and actual labels.

\section{Experiments}\label{sec:exp}
In this section, we perform experiments on two benchmark datasets and their variants to evaluate our proposed EDGMAT model. First, we present the setup of our experiments. Then, with both transductive and inductive settings on four datasets, we evaluate the performance of the proposed method under both static and dynamic topologies. Finally, we also compare our proposed method with state-of-the-art methods.

\subsection{Experimental Setup}

\subsubsection{Datasets}
To evaluate the proposed EDGMAT model, this study uses four publicly available network intrusion detection datasets in model training. These datasets have been widely used in many previous studies\cite{vinayakumar2020visualized,kumar2022p2tif,apruzzese2022cross}. They capture various network traffic characteristics, such as statistical packet characteristics, timing characteristics, and protocol flags. The Bot-IoT dataset is about IoT device traffic identification, which was used to study and evaluate malicious behaviors and abnormal activities in IoT devices. Published by Koroniotis et al. \cite{koroniotis2019towards} in 2019, this dataset contains traffic from IoT devices simulated in a laboratory environment. The ToN-IoT dataset is a new extensive network traffic dataset created by Alsaedi et al.\cite{alsaedi2020ton_iot} in 2020, which includes different types of IoT data. However, due to the current lack of standard formats, the huge differences in the features of various NIDS datasets make it difficult to compare the performance of ML-based network traffic classifiers on different datasets and evaluate their generalization capabilities to different network scenarios. Sarhan et al. \cite{sarhan2021netflow} addressed this issue by providing NetFlow versions of the two NIDS datasets mentioned above. The author used the captured packets in the .pcap format \footnote{https://fileinfo.com/extension/pcap} of the original NIDS dataset, converted it to NetFlow format through the nProbe tool \footnote{https://www.ntop.org/products/netflow/nprobe/}, and selected 12 fields for extraction, resulting in new variants of the original datasets, namely NF-ToN-IoT and NF -BoT-IoT. The NF-ToN-IoT dataset specifically captures traffic from IoT devices in a simulated laboratory environment. These datasets contain samples of normal and malicious traffic, including botnet attacks, and have been widely used to evaluate machine learning-based intrusion detection systems for IoT networks. Here, we provide a concise summary of these four datasets as follows:

\begin{itemize}
	\item[$\bullet$] \textbf{BoT-IoT}: The BoT-IoT dataset is for IoT device traffic identification, which can be used to study and evaluate malicious behavior and abnormal activity in IoT devices. This dataset focuses on the impact of malware and abnormal behavior on IoT networks. The authors used the Node-red tool \footnote{https://nodered.org/} to simulate various IoT devices, including hydropower stations, and generate the corresponding IoT traffic and then used the Argus tool \footnote{http://argus.tcp4me.com/} for feature extraction. The samples in the Bot-IoT dataset have a variety of characteristics, including source IP address, destination IP address, source port, destination port, transport protocol, and other traffic-related attributes. By analyzing these features, they can be used to train machine learning models to identify malicious behavior or unusual activity. The dataset is a 1.1 GB CSV file \footnote{Because training 1.1 GB textual data requires a huge amount of GPU memory, in our study we randomly extract 10\% data and keep the class distribution the same as the original one.}, consisting of six attack types with a total of 47 characteristics corresponding to type tags.
	\item[$\bullet$]\textbf{ToN-IoT}: This is a relatively new and extensive dataset, created by Alsaedi et al. \cite{alsaedi2020ton_iot} in 2020, which includes different types of IoT data, such as operating system logs, telemetry data from ToN-IoT services, and IoT network traffic collected from medium-size networks in the Cyber Range and IoT LABS in Canberra, New South Wales (Australia). The Bro-IDS network monitoring tool \footnote{https://old.zeek.org/manual/2.5.5/broids/index.html} was used to generate 44 network traffic characteristics of the dataset. In this study, we utilized the network traffic CSV dataset from the ToN-IoT dataset for training our model.
	\item[$\bullet$] \textbf{NF-ToN-IoT} and \textbf{NF-BoT-IoT}: The NF-ToN-IoT and NF-BoT-IoT datasets were created by Sarhan et al. \cite{sarhan2021netflow}. They retained and generated the common feature sets in the original two datasets ToN-IoT and NF-BoT-IoT, using the NetFlow format. The NF-ToN-IoT dataset consists of a total of 16,940,496 network flows. Among them, there are 10,841,027 attack samples, accounting for 63.99\%, and 6,099,469 benign samples, accounting for 36.01\%. This dataset encompasses nine attack categories, including Benign (36.01\%), Backdoor (0.10), DoS (4.21\%), DDoS (11.96\%), Injection (4.04\%), MITM (0.05\%), Password (6.89\%), Ransomware (0.02\%), Scanning (22.32\%), and XSS (14.49\%). The NF-BoT-IoT dataset comprises a total of 37,763,497 data streams, with 37,628,460 being attack samples (99.64\%) and 135,037 being benign samples (0.36\%). It includes four attack categories: Benign (0.36\%), Reconnaissance (6.94\%), DDoS (48.54\%), DoS (44.15\%), and Theft (0.01\%).
\end{itemize}

\subsubsection{Methods in Comparison}
To evaluate our proposed method, we compared it with several advanced methods currently available. We compared it with traditional machine learning methods that have already show their good performance on the four datasets as well as selected state-of-the-art graph neural network methods. The methods used in comparison are as follows:

\begin{itemize}
	\item[$\bullet$] \textbf{E-GraphSAGE}: Lo et al.\cite{Lo2021EGraphSAGEAG} proposed a graph neural network model named E-GraphSAGE for network intrusion detection, which is based on the traditional GraphSAGE method. E-GraphSAGE aggregates node information in network flow topology to generate higher-dimensional representations for determining the type of network flow attack. During the node aggregation process, this model utilizes neighborhood sampling and aggregation methods to aggregate the network flow features stored in the nodes. In this dynamic aggregation process, the network flow feature representations are continuously updated, and the classification results are obtained through perceptrons. This method is the first successful application of GNN in IoT network intrusion detection based on network flow data and has achieved high performance on multiple datasets.
	\item[$\bullet$] \textbf{KNN}:K-NearestNeighbor is a traditional supervised learning method. In order to explore the performance of various traditional machine learning methods in network intrusion detection, Churcher et al. \cite{churcher2021experimental} conducted a comprehensive empirical study, evaluating KNN, decision tree, support vector machine, Naive Bayes, random forest, artificial neural networks, and logistic regression on the above datasets. According to their experimental results, KNN classifier achieved the highest performance.
	\item[$\bullet$] \textbf{XGBoost}: XGBoost (eXtreme Gradient Boosting) is a powerful ensemble learning algorithm and a variant of Gradient Boosting Trees. It improves predictive performance gradually by iteratively training a series of decision tree models. Each new model attempts to correct the errors of the previous model, thereby gradually improving overall performance. Lawal et al. \cite{lawal2020anomaly} implemented a hybrid network intrusion detection system based on features and behaviors, using a database based on signatures and blacklist sources. The system employed XGBoost and achieved high accuracy in binary classification.
	\item[$\bullet$] \textbf{Extra-Trees Classifier}: Extremely Randomized Trees are an ensemble learning method based on random forests. At each node, it randomly selects a subset of features and a random split point within that subset for node splitting. Extra-Trees classifiers introduce additional randomness to enhance model diversity, mitigate the risk of overfitting, and enhance model robustness. Sarhan et al. \cite{sarhan2021netflow} has shown that Extra-Trees classifier achieves high performance on the above datasets.
\end{itemize}

\subsubsection{Settings of the Proposed EDGMAT model}
We used a simple four-layer GNN model. First, we applied two layers of EDGMAT convolutional layers that we defined  to map high-dimensional features. Then, the output values of the next layer were obtained through a ReLu activation function. The final layer is a softmax layer that calculates the probabilities of updated features for each class. In addition, in each convolutional layer, we defined a Dropout function with a value of 0.2 to prevent overfitting. In our experiments, we used the cross-entropy loss function to improve our network parameters, and we also incorporated class weights for each category in the loss function. We chose Adam as the optimizer with a learning rate of 0.01. The training iterations for all four datasets were mostly between 100 and 200, as we used a single-graph architecture to learn from all datasets in one round.

The machine that we run the experiments has a Silver 4210R @ 2.40GHz CPU and a RTX3080 10GB GPU. The programming languages and tools we used include Python, PyTorch, and DGL. In this paper, we used Precision, Recall, and F1-Score to quantify the performance of the model and analyze the classification results of each category in the datasets. Besides, to evaluate the overall performance of multi-class classification, we adopted weighted F1-Scores as a final metric, where weighted F1-Score $=\Sigma($ Class Weight $(C)$ * F1-Score $(C))$.

\subsection{Performance on the Static Topology}
Static topology network structure is linked with tranductive learning. Transductive learning is a more specific learning task that aims to make predictions for specific unknown data points rather than just generalizing to the entire data distribution. In our experiments, we used the entire data set to generate a complete and static topology map for subsequent performance testing of feature learning. Next, we specifically analyzed the classifier performance under this static topology (i.e., transductive setting).
\begin{table}[!t]
	\begin{center}
		\caption{Multi-class classification results of the proposed EDGMAT on dataset NF-BoT-IoT under the transductive setting}
		\label{tab2}
		\begin{tabular}{|c|c|c|c|}
			\hline    \multicolumn{4}{|c|}{Transductive}  \\
			\hline Class Name & Precision & Recall & F1-Score  \\
			\hline Benign & 22.99\% & 92.00\% & 0.37\\
			\hline DDos & 100.00\% & 91.54\% &0.96\\
			\hline Dos & 39\% & 97.79\% &0.56 \\
			\hline Reconnaissance & 99.90\% & 83.57\% & 0.91 \\
			\hline Theft & 21.23\% &99.68\% &0.35 \\
			\hline Weighted Average &95.88\% & 88.79\% & 0.91\\
			\hline
		\end{tabular}
	\end{center}
\end{table}

We explore the complexity of multi-class classification where classifiers are designed to distinguish between various types of attacks and benign traffic. This task presents a more difficult challenge than binary classification. We evaluate the performance of the proposed EDGMAT classifier in multi-class scenarios on above four datasets. The number of attack categories varies in these four datasets, ranging from four to nine. Table \ref{tab2} shows the experiment results on the  NF-BoT-IoT dataset when the network flow topology is constructed using a transductive setting. In the transductive setting, the entire network topological structure is ingested by the EDGMAT classifier. For each node, a sample of the test and verification set is used to gather information about its neighbor nodes. It is obvious that the EDGMAT classifier shows high accuracy, recall, and F1-scores when classifying DDoS and reconnaissance malicious traffic. It also has high performance in detecting malicious traffic in DDos and Reconnaissance. We observed that the classifier achieved good weighted accuracy (95.88\%) and weighted recall rate (88.79\%) and weighted F1-Scores (0.91) in the transductive setting.

\begin{table}[!t]
	\begin{center}
		\caption{Multi-class classification results of the proposed EDGMAT on dataset BoT-IoT under the transductive setting.}
		\label{tab3}
		\begin{tabular}{|c|c|c|c|}
			\hline    \multicolumn{4}{|c|}{Transductive}  \\
			\hline Class Name & Precision & Recall & F1-Score  \\
			\hline Benign & 84.65\% & 100.00\% & 0.92\\
			\hline DDos & 99.99\% &98.20\% &0.99\\
			\hline Dos & 97.97\% & 100.00\% & 0.99 \\
			\hline Reconnaissance & 99.90\% & 99.82\% & 1.00 \\
			\hline Theft & 0\% & 0\%  &  0 \\
			\hline Weighted Average & 99.94\%  & 99.98\% & 0.99\\
			\hline
		\end{tabular}
	\end{center}
\end{table}

The EDGMAT classifier, when configured with the transductive setting, exhibited outstanding performance on the BoT-IoT dataset. Notably, it achieved an exceptional W-F1 score of 0.99, as evident in the outcomes displayed in Table \ref{tab3}. An examination of the dataset's structure reveals a minimal representation of benign traffic in comparison to malicious traffic, particularly within the theft category. The dataset predominantly comprises the other three types of malicious traffic, which contributed to the classifier's inability to effectively identify these specific categories.
\begin{table}[!t]
	\begin{center}
		\caption{Multi-class classification results of the proposed EDGMAT on dataset NF-ToN-IoT under the transductive setting.}
		\label{tab4}
		\begin{tabular}{|c|c|c|c|}
			\hline    \multicolumn{4}{|c|}{Transductive}  \\
			\hline Class Name & Precision & Recall & F1-Score  \\
			\hline Benign & 76.80\% & 95.00\% & 0.85\\
			\hline Backdoor & 100.00\% & 99.80\%  & 0.99\\
			\hline DDos & 80.00\% & 58.92\% & 0.68\\
			\hline Dos & 29.70\% & 100\% &0.46 \\
			\hline Injection & 99.10\%  & 10.60\% & 0.19\\
			\hline MIMT & 5.31\%  & 46\%  &0.10 \\
			\hline Password & 26.43\% & 47.26\% & 0.34\\
			\hline Ransomware & 3.00\% & 100.00\% & 0.06 \\
			\hline Scanning & 7.50\% & 15.74\% & 0.10\\
			\hline XSS & 15.00\% & 62.14\% & 0.24 \\
			\hline Weighted Average & 74.41\% & 67.75\% & 0.74\\
			\hline 
		\end{tabular}
	\end{center}
\end{table}

On the NF-ToN-IoT dataset, we employed the EDGMAT algorithm with the transductive setting to assess the model's performance. As illustrated in Table \ref{tab4}, the classifier achieved a weighted F1-Score of 0.74 under the transductive setting. A thorough analysis indicates that the classifier excels in categories characterized by a substantial volume of network flows, such as the Backdoor category. However, it exhibits suboptimal performance in categories marked by a limited number of network flows, such as the Dos category.

\begin{table}[!t]
	\begin{center}
		\caption{Multi-class classification results of the proposed EDGMAT on dataset ToN-IoT under the transductive setting.}
		\label{tab5}
		\begin{tabular}{|c|c|c|c|}
			\hline    \multicolumn{4}{|c|}{Transductive}  \\
			\hline Class Name & Precision & Recall & F1-Score  \\
			\hline Backdoor & 87.30\% & 99.90\% & 0.93\\
			\hline DDos & 91.90\% & 81.71\%  & 0.89\\
			\hline Dos & 85.00\% & 94.21\% & 0.89\\
			\hline Injection & 68.32\% & 63.10\% & 0.60\\
			\hline MIMT & 67.81\% & 34.20\% & 0.46\\
			\hline Normal &94.90\% &98.71\% & 0.97 \\
			\hline Password & 62.70\% & 87.90\%& 0.73\\
			\hline Ransomware & 96.78\%& 98.00\% & 0.97\\
			\hline Scanning & 91.47\% & 21.64\% & 0.35\\
			\hline XSS & 90.69\% &73.08\% & 0.81\\
			\hline Weighted Average & 91.62\% & 86.40\% & 0.91\\
			\hline 
		\end{tabular}
	\end{center}
\end{table}

As can be seen from Table \ref{tab5}, the EDGMAT classifier has good results when applied to ToN-IoT datasets. Under the transductive setting, the EDGMAT classifier shows good accuracy in various classes with small data volumes, such as 68.32\% in the Injection class and 67.81\% in the MIMT class. A high F1-Score was achieved in both categories, and the overall weighted F1-Scores reached 0.92.

\begin{figure}[htbp]
	\centering
	\subfloat[Raw data]
	{
		\includegraphics[width=0.22\textwidth]{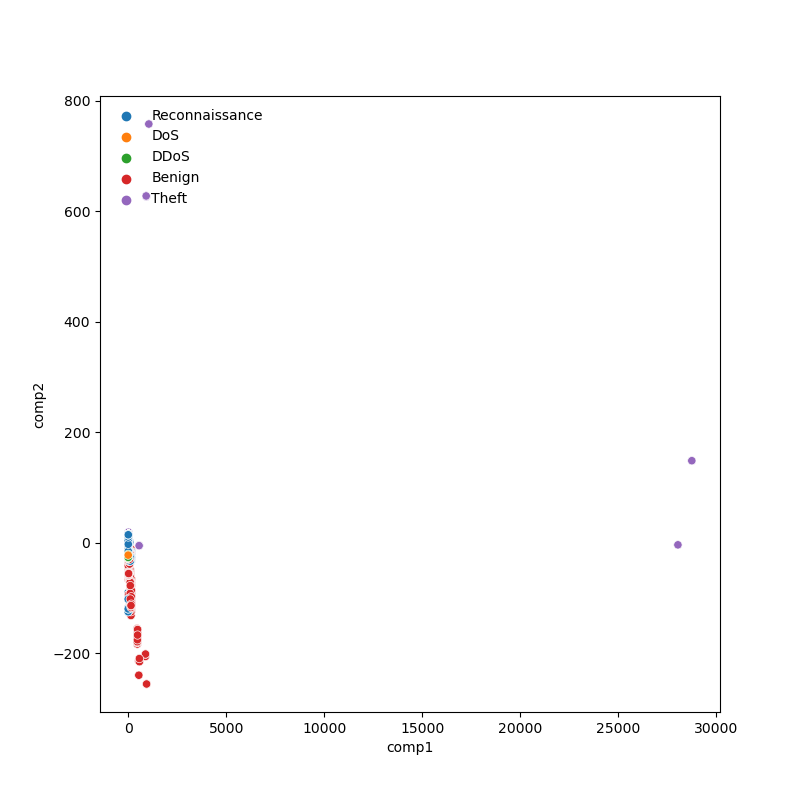}
	}
	\quad
	\subfloat[Edge embedding after convolution]
	{
		\includegraphics[width=0.22\textwidth]{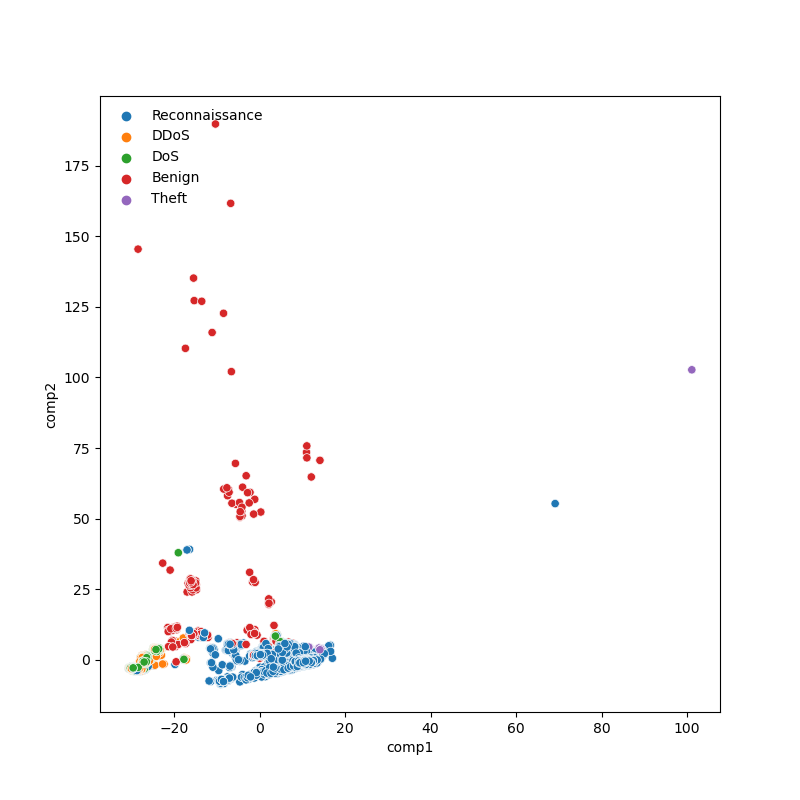}
	}
	\caption{Dimensionality reduction visualization using PCA a) 10\% raw samples of NF-BoT-IoT, b) EDGMAT-generated edge embedding samples}
	\label{fig3}
\end{figure}
\begin{figure}[htbp]
	\centering
	\subfloat[Raw data]
	{
		\includegraphics[width=0.22\textwidth]{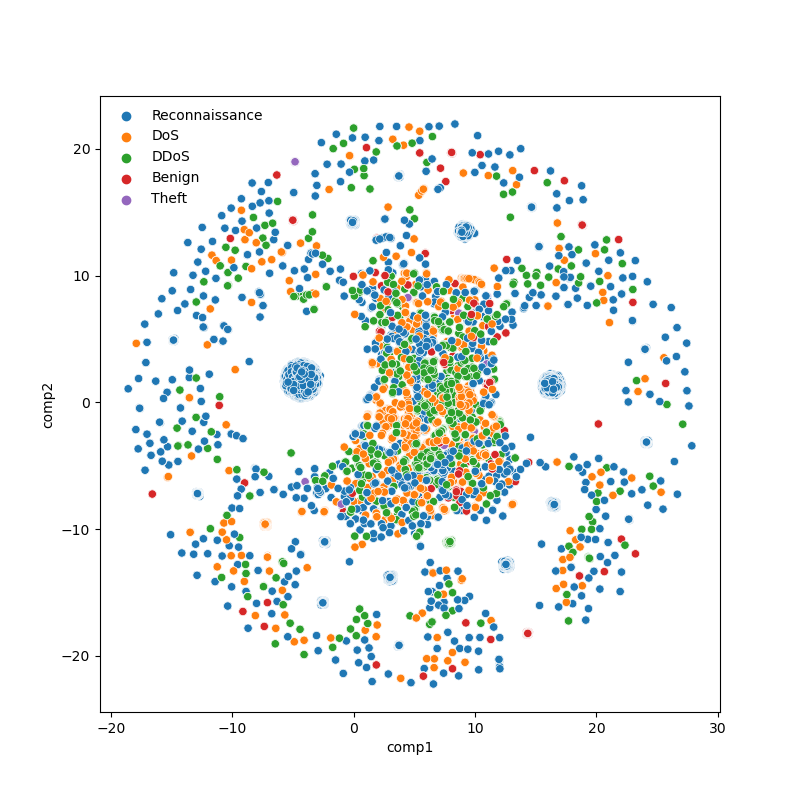}
	}
	\quad
	\subfloat[Edge embedding after convolution]
	{
		\includegraphics[width=0.22\textwidth]{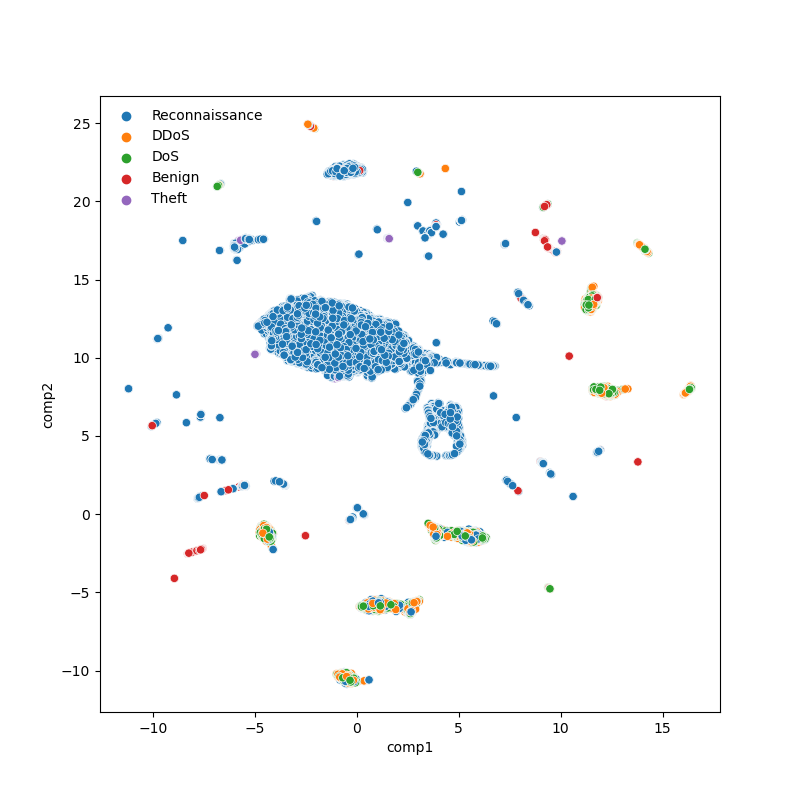}
	}
	\caption{Dimensionality reduction visualization using UMAP a) 10\% raw samples of NF-BoT-IoT, b) EDGMAT-generated edge embedding samples}
	\label{fig4}
\end{figure}

In order to gain a clear understanding of the classification ability of our proposed EDGMAT classifier, we have investigated and visualized the edge embeddings generated by the algorithm on 10\% samples of the NF-BoT-IoT dataset. The original features and the edge features obtained through convolution operation are mapped from high-dimensional data to two-dimensional data using UMAP and PCA algorithms for visualization. As shown in Figure \ref{fig3} and \ref{fig4}, we can observe that networks of the same type aggregate together, while networks of different types are distinguished by their different feature values. This indicates that our EDGMAT classifier is capable of classifying different types of malicious traffic and benign traffic based on the topological model of network flows and traffic feature values.

\subsection{Performance on the Dynamic Topology}
Dynamic topology network structure is associated with inductive learning, which represents a prevalent machine learning approach for deriving overarching principles or patterns from a limited training dataset. This facilitates the extrapolation of unseen data. In the inductive learning paradigm, the model scrutinizes and incorporates the attributes and associations within the training data, ultimately building a generalized model to predict and classify unobserved data.

For our specific task, we make use of the data to construct a dynamic topology during the training phase to facilitate parameter learning. In the subsequent testing phase, we employed the complete test dataset to create a fresh topology. Within the inductive setting, we partitioned the NIDS dataset into a training set and a test set. The training set is utilized to build a topology diagram for training, while the test set is employed to generate a distinct topology diagram specifically for network traffic classification during testing.

Furthermore, a new topology diagram is dynamically generated based on the available data and network addresses within each application. In this approach, the EDGMAT model is exclusively trained on the training samples, which enables the processing of new graph nodes and the generation of embeddings for unknown nodes by leveraging information derived from known nodes.

\begin{table}[!t]
	\begin{center}
		\caption{Multi-class classification results of the proposed EDGMAT on dataset NF-BoT-IoT under the inductive setting.}
		\label{tab6}
		\begin{tabular}{|c|c|c|c|}
			\hline    \multicolumn{4}{|c|}{Inductive}  \\
			\hline Class Name & Precision & Recall & F1-Score  \\
			\hline Benign & 26.73\% & 92.18\% & 0.41\\
			\hline DDos & 100.00\% & 91.70\% &0.96\\
			\hline Dos & 39.13\% & 97.71\% &0.56 \\
			\hline Reconnaissance & 99.68\% & 85.08\% & 0.92 \\
			\hline Theft & 16.06\% &74.35\% &0.26 \\
			\hline Weighted Average & 96.11\%  & 89.41\% & 0.92\\
			\hline
		\end{tabular}
	\end{center}
\end{table}
Table \ref{tab6} presents the results obtained through the inductive setting's approach to graph construction applied to the NF-BoT-IoT dataset. When utilizing the NetFlow format feature set within the dataset, the EDGMAT classifier exhibits elevated levels of accuracy, recall, and F1-scores in the classification of DDoS, Dos, and Reconnaissance traffic. It is noteworthy that, within the inductive setting, the classifier achieves superior weighted accuracy (96.11\%), weighted recall rate (89.41\%), and weighted F1-Score (0.92) when compared to the transductive setting

\begin{table}[!t]
	\begin{center}
		\caption{Multi-class classification results of the proposed EDGMAT on dataset BoT-IoT under the inductive setting.}
		\label{tab7}
		\begin{tabular}{|c|c|c|c|}
			\hline    \multicolumn{4}{|c|}{Inductive}  \\
			\hline Class Name & Precision & Recall & F1-Score  \\
			\hline Benign & 84.75\% & 93.25\% & 0.89\\
			\hline DDos & 99.99\% &98.54\% &0.99\\
			\hline Dos & 99.97\% & 99.99\% & 0.99 \\
			\hline Reconnaissance & 99.99\% & 99.71\% & 1.00 \\
			\hline Theft & 0\% & 0\%  &  0 \\
			\hline Weighted Average & 99.99\%  & 99.98\% & 0.99\\
			\hline
		\end{tabular}
	\end{center}
\end{table}

Table \ref{tab7} illustrates the remarkable performance of the classifier on the BoT-IoT dataset under the inductive setting. Moreover, the classifier's performance within each specific category consistently mirrors the performance achieved in the transductive setting.

\begin{table}[!t]
	\begin{center}
		\caption{Multi-class classification results of the proposed EDGMAT on dataset NF-ToN-IoT under the inductive setting.}
		\label{tab8}
		\begin{tabular}{|c|c|c|c|}
			\hline   \multicolumn{4}{|c|}{Inductive}  \\
			\hline Class Name & Precision & Recall & F1-Score  \\
			\hline Benign & 95.58\% & 83.83\% & 0.89\\
			\hline Backdoor & 99.59\% & 98.83\%  & 0.99\\
			\hline DDos & 0\% & 0\% & 0\\
			\hline Dos & 0\% & 0\% &0 \\
			\hline Injection & 77.69\%  & 20.22\% & 0.32\\
			\hline MIMT & 1.32\%  & 90.72\%  &0.03 \\
			\hline Password & 0\% & 0\% & 0\\
			\hline Ransomware & 0.06\% & 97.67\% & 0.01 \\
			\hline Scanning & 3.14\% & 47.42\% & 0.06\\
			\hline XSS & 16.84\% & 62.52\% & 0.27 \\
			\hline Weighted Average & 74.41\% &65.42\%& 0.67\\
			\hline
		\end{tabular}
	\end{center}
\end{table}

On the NF-ToN-IoT dataset, as depicted in Table \ref{tab8}, it is evident that the performance of the classifier is notably inferior in dynamic topology when contrasted with the static topology. A more detailed analysis reveals that, within categories featuring a substantial number of network flows, the classifier excels in the inductive setting. However, it struggles, even achieving an F1-Score of 0, in categories with fewer network flows. On the contrary, the classifier in the transductive setting also exhibits proficiency in categories with limited data instances, notably achieving a 100\% recall rate in the Dos and ransomware categories. The disparities between the EDGMAT classifiers under distinct settings are discernible. The transductive setting enables the EDGMAT classifier to manage imbalanced datasets and offers some capability to recognize categories with fewer data instances. Conversely, the EDGMAT classifier in the inductive setting lacks the ability to discern imbalanced categories but excels in identifying malicious traffic when an adequate sample size is available. Furthermore, as it does not necessitate all the topology patterns, known nodes can be employed to generate unknown nodes, indicating a robust generalizability of this approach.
\begin{table}[!t]
	\begin{center}
		\caption{Multi-class classification results of the proposed EDGMAT on dataset ToN-IoT under the Inductive setting.}
		\label{tab9}
		\begin{tabular}{|c|c|c|c|}
			\hline    \multicolumn{4}{|c|}{Inductive}  \\
			\hline Class Name & Precision & Recall & F1-Score  \\
			\hline Backdoor &99.95\% & 99.40\% & 0.99\\
			\hline DDos & 97.75\% & 96.29\%  & 0.97\\
			\hline Dos & 91.23\% & 84.60\% & 0.88\\
			\hline Injection & 13.03\% & 14.10\% & 0.13\\
			\hline MIMT & 10.75\% & 91.53\% & 0.19\\
			\hline Normal &99.91\% &87.68\% & 0.93 \\
			\hline Password & 44.81\% & 86.07\%& 0.59\\
			\hline Ransomware & 76.35\%& 94.73\% & 0.85\\
			\hline Scanning & 71.61\% & 93.87\% & 0.81\\
			\hline XSS & 88.25\% &64.57\% & 0.75\\
			\hline Weighted Average & 93.78\% & 90.89\% & 0.92\\
			\hline 
		\end{tabular}
	\end{center}
\end{table}

Table \ref{tab9} exhibits the results obtained through the graph construction method of the inductive setting within the ToN-IoT dataset. It was observed that under the inductive setting, as compared to the transductive setting, the classifier achieved notably superior performance with higher weighted accuracy (93.78\%), weighted recall rate (90.89\%), and a weighted F1-score of 0.92. Nonetheless, in the ToN-IoT dataset, the classifier under the inductive setting still demonstrates suboptimal performance in two categories, namely Injection and MIMT. Nevertheless, due to their smaller dataset sizes, these categories do not exert a substantial influence on the final weighted results, and the classifier exhibits improved performance under the dynamic topology setting.

\subsection{Comparison with State-Of-The-Art Methods}
In our study, we conducted experiments to obtain multiple classification results using the EDGMAT classifier on the above four distinct datasets. These results are subsequently compared with the state-of-the-art methods currently employed in the field of network intrusion detection.

\begin{table}[!t]
	\begin{center}
		\caption{Performance of multiclass classification by EDGMAT compared with the state-of-art algorithms under the static topology.}
		\label{tab10}
		\begin{tabular}{|c|c|c|}
			\hline  Method & Dataset & W-F1 \\
			\hline \textbf{EDGMAT}  & BoT-IoT & \textbf{0.99}\\
			 KNN & BoT-IoT &0.99 \\
			 XGBoost & BoT-IoT & 0.97 \\
			\hline \textbf{EDGMAT}  & NF-BoT-IoT & \textbf{0.91} \\
			KNN & NF-BoT-IoT & 0.80 \\
			XGBoost & NF-BoT-IoT & 0.83 \\
			 Extra Tree Classifier & NF-BoT-IoT & 0.77\\
			\hline \textbf{EDGMAT}  & NF-ToN-IoT  & \textbf{0.74}  \\
			KNN  & NF-ToN-IoT & 0.61 \\
			XGBoost & NF-ToN-IoT & 0.60 \\
			 Extra Tree Classifier & NF-ToN-IoT & 0.60\\
			\hline \textbf{EDGMAT}  &ToN-IoT & \textbf{0.91}\\
			KNN & ToN-IoT & 0.97 \\
			XGBoost & ToN-IoT & 0.88 \\
			 Extra Tree Classifier & ToN-IoT & 0.87\\
		   	\hline
		\end{tabular}
	\end{center}
\end{table}
Table \ref{tab10} provides a comparison between the EDGMAT model proposed in this paper under static topology and the best results reported in recent literature. We observe that the performance of the EDGMAT classifier surpasses that of state-of-the-art classifiers on three datasets. Specifically, on the BoT-IoT dataset, our method and the KNN approach both achieved an optimal result of 0.99. In contrast, the KNN method outperformed on the ToN-IoT dataset, achieving a weighted F1-Score of 0.97. This can be attributed to our selection of a specific subset of the dataset, preprocessed by the authors, for training and testing purposes. This subset exhibits a balanced distribution of different classes and uniform data feature distribution, making it particularly well-suited for the KNN method. Overall, our approach employs graph neural networks, offering advantages in terms of generalization and performance in handling sparse data. Additionally, our method, in the context of static topology, generally outperforms most of the current machine learning-based network intrusion detection methods.

\begin{table}[!t]
	\begin{center}
		\caption{Performance of multiclass classification by EDGMAT compared with state-of-art algorithms under the dynamic topology.}
		\label{tab11}
		\begin{tabular}{|c|c|c|}
			\hline  Method & Dataset & W-F1 \\
			\hline \textbf{EDGMAT} &BoT-IoT &  \textbf{0.99} \\
			
			 E-GraphSAGE & BoT-IoT & 0.99\\
			
			\hline \textbf{EDGMAT}  & NF-BoT-IoT & \textbf{0.92}\\
		
			 E-GraphSAGE &NF-BoT-IoT  & 0.81 \\

			\hline \textbf{EDGMAT}  & NF-ToN-IoT & \textbf{0.67}\\
			E-GraphSAGE & NF-ToN-IoT &0.63  \\
		
			\hline \textbf{EDGMAT}  & ToN-IoT & \textbf{0.92} \\
		
			E-GraphSAGE & ToN-IoT & 0.87  \\
			
			\hline
		\end{tabular}
	\end{center}
\end{table}
Since E-GraphSAGE can handle dynamic topology, Table \ref{tab11} presents a performance comparison between the EDGMAT method and the E-GraphSAGE method in such an environment. It can be observed that both methods achieved a weighted F1-score of 0.99 on the BoT-IoT dataset. However, our proposed method outperformed E-GraphSAGE on the other three datasets. These comparative experiments demonstrate the excellent performance of our proposed EDGMAT method.

\section{Conclusion}\label{sec:con}
This paper proposes EDGMAT, a network intrusion detection method founded on the principles of a graph attention network. The approach harmonizes network flow attributes with the topological model of networks, encompassing the directional aspects of network traffic and imposing distinct weights on neighboring nodes during the aggregation phase. It exhibits an adept capability to effectively discern various forms of malicious traffic. This study primarily focuses on the incorporation of traffic directionality within network topology, alongside the utilization of a multi-head attention mechanism in the domain of network intrusion detection. Furthermore, it takes into account the nuanced performance nuances of the model in the context of dynamic and static topologies. In this research endeavor, we realized a network intrusion detection model grounded in EDGMAT and executed experiments across four extensively employed datasets in the realm of Network Intrusion Detection Systems (NIDS). The empirical assessments conducted on these datasets conclusively underline the model's formidable proficiency, underscoring the auspicious potential of graph attention networks within the domain of intrusion detection. Additionally, our proposed methodology necessitates further scrutiny of the computational and temporal overhead associated with the model. Mitigating GPU memory usage and training duration while upholding a high level of accuracy presents a prime avenue for prospective enhancement.



%

\bibliographystyle{IEEEtran}
\bibliography{ref}

\vfill

\end{document}